\documentclass[aps,twocolumn,10pt,pra,floatfix,nofootinbib, superscriptaddress]{revtex4-2}
\usepackage{siunitx}
\usepackage{amsmath}
\usepackage{amssymb}
\usepackage{bm}
\usepackage[retainorgcmds]{IEEEtrantools}
\usepackage{graphicx}
\usepackage{mathrsfs}
\usepackage{color}
\usepackage{amsfonts}
\usepackage{mhchem}
\usepackage{tikz}
\usepackage[colorlinks=true,linkcolor=rot,urlcolor=rot,citecolor=rot]{hyperref}
\usepackage{lipsum}
\usepackage{soul}
\usepackage{physics}
\usepackage{float}
\usepackage{dsfont}
\usepackage{mathrsfs}
\usepackage{BOONDOX-ds}
\usepackage{marvosym}
\usepackage{nicefrac}
\usepackage{bbold}
\usepackage{changepage}

\definecolor{mint}{RGB}{164, 215, 210} \definecolor{rot}{RGB}{210, 5, 55}
\definecolor{lightrot}{RGB}{235, 130, 155} \definecolor{antri}{RGB}{45, 55, 60}
\definecolor{RoyalBlue}{cmyk}{1, 0.50, 0, 0}

\begin{document}

\title{Autonomous optomechanical  Maxwell demon engine}

\author{Sander Stammbach}
\email{sander.stammbach@uni-siegen.de}
\author{Giulio Gasbarri}
\author{Stefan Nimmrichter}
\affiliation{Naturwissenschaftlich–Technische Fakultät, Universität Siegen,
    \\Walter-Flex-Straße 3, 57068 Siegen, Germany }

\date{\today}% It is always \today, today,
\begin{abstract}
We present a fully autonomous model of a mechanical Maxwell demon engine that converts heat absorbed by a qubit into ergotropy in an attached quantum battery. A damped mechanical oscillator acts as a pointer that incorporates the essential demon function of information acquisition about the qubit state and subsequent feedback into the system dynamics, replacing external measurements and control operations. This is achieved by means of a Holstein-type Hamiltonian, which makes the pointer’s equilibrium position depend on the qubit state and thereby induces a state-dependent resonance shift. The mechanism enables a selective energy transfer to an attached spin system that acts as the quantum battery. We derive both local and global Lindblad master equations for the coupled system and identify the parameter regimes in which the engine operates efficiently. The figure of merit is the energy stored in the battery, quantified by its ergotropy. Based on these results, we discuss the necessary conditions for practical implementations of the demon engine on realistic experimental platforms.
\end{abstract}

\maketitle

\section{Introduction}

Quantum heat engines offer a platform to explore the interplay between thermodynamics and quantum mechanics. They reveal how quantum effects such as coherence and entanglement can influence the conversion of heat into work \cite{scovil_three-level_1959,Friedenberger_2017}.
A paradigmatic example is the Szilard engine~\cite{Szilard1929,maruyama_colloquium_2009}, inspired by Maxwell's famous thought experiment. There, a ``demon'' uses information about microscopic fluctuations to extract useful work, seemingly in conflict with the second law of thermodynamics. Modern treatments resolve this paradox by accounting for the thermodynamic role of information processing and memory erasure~\cite{landauer_irreversibility_1961,bennett_notes_2003,PhysRevE.85.021104,deOliveiraJunior2025}.

Measurement-based implementations of Maxwell's demon have been demonstrated on several platforms~\cite{Goerlich_2025,deOliveiraJunior2025}, including photonic systems~\cite{PhysRevLett.116.050401}, superconducting circuits~\cite{Koski2015,masuyama_information-work_2018,
PhysRevLett.121.030604}, and quantum dot systems \cite{Barker2022}. In such realizations, an external controller typically measures the working medium and applies feedback conditioned on the measurement outcome~\cite{Elouard2017,ElouardHerreraHuardAuffeves2017,PhysRevLett.121.030604}. A complete thermodynamic description must therefore also account for the physical resources required to operate the measurement-and-feedback apparatus~\cite{alicki2013modelquantummeasurementthermodynamical,Alicki_2019,faist_minimal_2015}.
Recent work on information engines has moved beyond proof-of-principle demonstrations towards questions of performance, finite-time operation, and increasingly complex nonequilibrium working media~\cite{Goerlich_2025}.

More recent approaches have explored autonomous and semi-autonomous quantum demon engines, in which the measurement-and-feedback is reproduced by internal degrees of freedom and their interactions, without requiring an external agent to perform a prescribed sequence of measurements and control operations.
Examples include superconducting circuits, where information about a thermally excited qubit is encoded in a dispersively coupled microwave cavity and used to enable selective energy extraction~\cite{cottet:hal-01626961}; Rydberg atoms in microwave cavities, where an auxiliary atomic state controls excitation transfer~\cite{PhysRevResearch.2.032025}; and an autonomous dissipative Maxwell demon realized with a nitrogen-vacancy centre in diamond~\cite{Hern_ndez_G_mez_2022}. Related demon schemes have also been proposed for coupled quantum dots~\cite{LucasTrigal2025,Monsel2025,Ptaszynski2018}.

In fully autonomous settings, the operational roles of measurement, memory, and feedback are encoded in the system dynamics, rather than implemented as separate control steps. Such engines may therefore be viewed as self-oscillating quantum thermal devices that convert a temperature bias into useful, accumulated energy~\cite{Alicki2021engines}.
Still, autonomous demon engines retain the measurement-and-feedback logic of a Maxwell demon, even though these operations are not performed explicitly. This distinguishes them conceptually from conventional continuous or cyclic heat engines, whose operation need not rely on an information-conditioned mechanism.

Building on previous work~\cite{Seah2020}, we introduce an autonomous Maxwell-demon engine in which the working medium is a two-level system (qubit) permanently coupled to a hot thermal bath. A damped harmonic oscillator acts as a mechanical pointer whose equilibrium position depends on the qubit state. This state dependence is generated by a Holstein-type interaction, i.e. excitation of the qubit displaces the oscillator equilibrium position. Provided that the pointer relaxes sufficiently rapidly compared with the qubit dynamics, its position tracks the qubit state and thereby becomes correlated with it.

The same qubit--pointer coupling also makes the qubit transition frequency depend on the pointer position. As the pointer relaxes towards the position associated with the excited qubit state, the resulting frequency shift brings the qubit into resonance with an auxiliary multilevel spin system that serves as a quantum battery.
Excitation transfer to the battery is therefore selectively enhanced when the qubit is excited and suppressed otherwise. The pointer thereby reproduces the operational structure of a Maxwell demon. It acquires information about the qubit state through its displacement and subsequently conditions the energy transfer to the battery.

The pointer relaxation is provided by coupling the oscillator to a cold thermal bath.
The resulting mechanism therefore operates without explicit measurements or time-dependent feedback control.
Since the device is driven by a temperature difference between an hot and cold reservoir, it can also be viewed as a continuously operating thermal machine that converts a temperature bias into non-passive energy stored in the battery.

The qubit--pointer interaction is described by the Holstein Hamiltonian, originally introduced to describe electron--phonon coupling and polaron formation in molecular crystals~\cite{holstein_polaron_1959}. It has since been widely used in vibronic and optovibronic systems, where electronic transitions couple to molecular vibrations and, in some cases, to optical cavities~\cite{PhysRevLett.122.203602,Ooi2022,zeb_exact_2018}. Holstein-type interactions also arise in opto- and electromechanical platforms~\cite{PhysRevB.80.155411}. These systems can support strong qubit--oscillator couplings and additional degrees of freedom for energy storage, making them promising candidates for realizing the mechanism considered here.

The article is organized as follows. In Sec.~\ref{sec:model}, we introduce the autonomous qubit--pointer--battery model and its open-system dynamics.
In Sec.~\ref{sec:performance}, we analyze the steady-state charging performance and quantify the useful energy stored in the battery in terms of population inversion and ergotropy~\cite{Allahverdyan2004}.
In Sec.~\ref{glob_vs_loc}, we compare the local Lindblad description with global Born--Markov--secular (BMS) and second-order time-convolutionless (TCL2) expansions~\cite{BreuerPetruccione2002,Schaller2014}. In Sec.~\ref{sec:platforms}, we discuss possible experimental implementations, before concluding in Sec.~\ref{sec:conclusion}.

\section{Model}\label{sec:model}

\begin{figure}[h]
    \centering
    \includegraphics[width=1.15\linewidth]{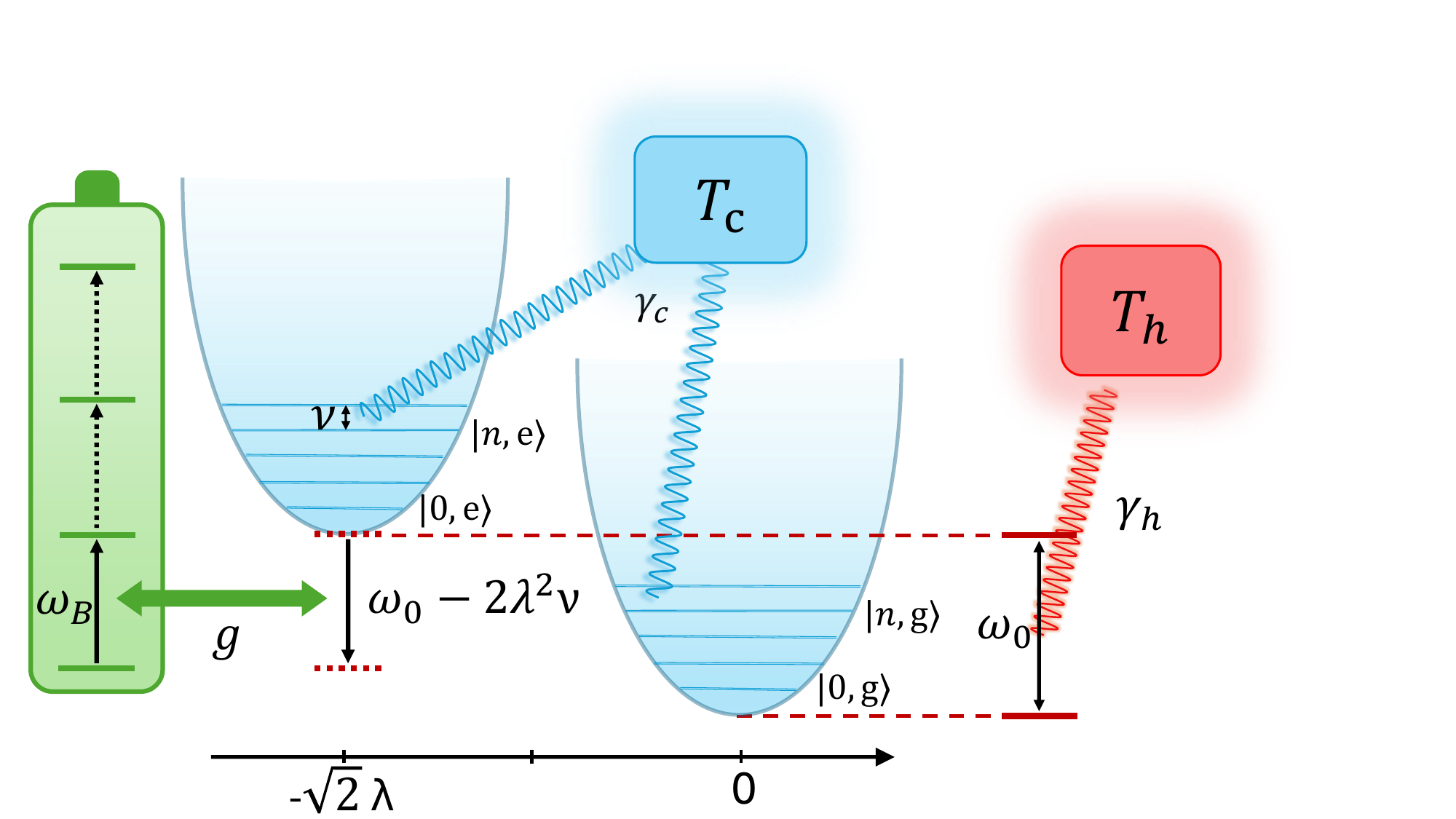}
    \caption{Schematic representation of the autonomous demon engine. A qubit (red) with bare transition frequency $\omega_0$ is coupled to a harmonic-oscillator pointer (blue) of frequency $\nu$. The qubit is coupled to a hot bath at temperature $T_h$ with rate parameter $\gamma_h$. Depending on the qubit state, the pointer potential is centered at $x_g=0$ or $x_e=-\sqrt{2}\lambda$. A cold bath at temperature $T_c$ damps the pointer at rate $\gamma_c$. The green multilevel system represents the battery with level spacing $\omega_B$, exchange-coupled to the qubit with strength $g$.}
    \label{model}
\end{figure}

We consider an open quantum system composed of three interacting subsystems: a two-level working medium (S), a harmonic-oscillator pointer (O), and a multilevel quantum battery (B), as illustrated in Fig.~\ref{model}. Throughout, we set $\hbar=k_{\rm B}=1$, such that energies and temperatures are expressed in units of angular frequency.

The working medium is a two-level system, with bare transition frequency $\omega_0$, weakly coupled to a hot thermal bath at temperature $T_h$ with characteristic coupling-rate parameter $\gamma_h\ll\omega_0$. The pointer is a mechanical oscillator of frequency $\nu\ll\omega_0$, whose equilibrium position depends on the qubit state. It is damped by a cold bath at temperature $T_c$ with rate $\gamma_c$. We focus on regimes in which the pointer relaxes sufficiently rapidly compared with the characteristic qubit-state switching times and in which the two conditional pointer states are sufficiently well separated to encode the qubit state. At low pointer temperature, this requires a displacement parameter of order $\lambda\gtrsim1$.

The coupled qubit--oscillator subsystem is described by the Holstein Hamiltonian
\begin{align}
  \hat{H}_{\mathrm{SO}}
    &= \omega_0 \hat{\sigma}^\dagger \hat{\sigma}
    + \lambda \nu (\hat{b}^\dagger + \hat{b}) \hat{\sigma}^\dagger \hat{\sigma}
    + \nu \hat{b}^\dagger \hat{b}
    \nonumber \\ %\label{eq:HSO} \\
    &= \left( \omega_0 + \nu x_0 \hat{x} \right) \hat{\sigma}^\dagger \hat{\sigma}+\frac{\nu}{2} (\hat x^2 + \hat p^2) - \frac{\nu}{2}
    .\label{eq:HSO}
\end{align}
where $\lambda$ is the dimensionless qubit--pointer coupling strength, $x_0=\sqrt{2}\lambda$. Here, $\hat\sigma^\dagger=\ket{e}\!\bra{g}$ and $\hat{\sigma}=\ket{g}\!\bra{e}$ are the qubit raising and lowering operators, and $\hat{b}^{\dagger}$ and $\hat{b}$ are the creation and annihilation operators of the pointer. In the second line, we introduced the dimensionless quadratures $\hat x= (\hat b + \hat b^{\dagger})/\sqrt{2}$ and $\hat p = (\hat b - \hat b^{\dagger})/\sqrt{2}i$.

Equation~\eqref{eq:HSO} reveals the two roles played by the qubit--pointer interaction. First, it produces a qubit-state-dependent force on the oscillator. For a qubit in the ground state $\ket{g}$, the oscillator potential is centered at $x_{g}=0$, whereas in the excited state $\ket{e}$, it is centered at $x_{e}=-x_{0}=-\sqrt{2}\lambda$.
The pointer therefore becomes correlated with the qubit state as it relaxes towards the corresponding conditional equilibrium position.
The same coupling makes the qubit resonance frequency conditional on the pointer coordinate. For any fixed value $x$ of the pointer coordinate, Eq.~\eqref{eq:HSO} gives the corresponding qubit transition frequency
$\omega_{\rm{S}}(x)=\omega_0+\sqrt{2}\lambda\nu x$.
This quantity should be distinguished from the transition frequencies between eigenstates of the fully coupled qubit--oscillator Hamiltonian, which form a vibronic sideband structure~\cite{Franck1924,holstein_polaron_1959}; see also Appendix~\ref{app:FC_sidebands}.
Nevertheless, it provides a useful intuitive picture: as the pointer moves towards the displaced equilibrium associated with the excited qubit state, the relevant transition energies become increasingly red-detuned.

To extract energy selectively from the qubit, we couple it to an auxiliary multilevel system that acts as a quantum battery (B).
The battery is modeled as a spin-$j$ system with half-integer quantum number $j$ and canonical spin operators $(\hat{J}_x,\hat{J}_y,\hat{J}_z)$, spanned by the $n=2j+1$ eigenstates $\ket{m}$ associated to the $\hat{J}_z$-eigenvalues $m\in\{-j,-j+1,\ldots, j\}$. For the bare battery Hamiltonian, we use
\begin{equation}
    \hat{H}_{\mathrm{B}}=\omega_B\left(\hat{J}_{z}+j\mathbb{1}\right),
    \label{eq:HB}
\end{equation}
which describes $n$ equally spaced energy levels separated by $\omega_{\rm B}$, from zero to $(n-1)\omega_B$. 
The qubit and battery exchange single excitations through
\begin{equation}
    \hat{H}_{\mathrm{SB}}=g\left(   \hat{\sigma}\hat{J}_{+}+\hat{\sigma}^{\dagger}\hat{J}_{-}\right),
    \label{eq:HSB}
\end{equation}
where $\hat{J}_{\pm}=\hat{J}_{x}\pm i\hat{J}_{y}$ are the standard spin ladder operators~\cite{sakurai}. The term $\hat{\sigma}\hat{J}_{+}$ transfers one excitation from the qubit to the battery, whereas $\hat{\sigma}^\dagger\hat{J}_{-}$ describes the reverse process.

This exchange is resonantly enhanced when the battery transition frequency matches a relevant transition of the coupled qubit--pointer system and is suppressed when the corresponding detuning is large compared with the effective exchange coupling and transition linewidth. We therefore choose a red-detuned battery frequency, $\omega_{B}<\omega_{0}$, such that the exchange becomes favorable after the qubit has been excited and the pointer has moved towards its displaced equilibrium position.

Upon thermal excitation of the qubit, the pointer relaxes towards the displaced equilibrium position $\langle\hat x\rangle\approx-x_0$, corresponding to a conditional transition energy  $  \omega_{\rm{S}}(-x_0)= \omega_0-2\lambda^2\nu$.
This provides an estimate for the characteristic battery resonance. We expect efficient state-selective excitation transfer when the battery is tuned close to, or somewhat below, this frequency, $\omega_B \lesssim \omega_{\rm S}(-x_0)  = \omega_0-2\lambda^2\nu$.
In this regime, the battery interacts preferentially with the qubit after excitation and pointer relaxation, allowing it to accumulate useful energy in the form of a steady-state population inversion. As shown in Appendix~\ref{app:FC_sidebands}, the characteristic frequency $\omega_0-2\lambda^2\nu$ also corresponds to the dominant Franck--Condon sideband in the low-temperature, sufficiently large-displacement regime. The broader operation range towards lower battery frequencies is investigated numerically in Sec.~\ref{sec:performance}.

The operating mechanism can therefore be summarized as follows. A thermal excitation from the hot bath promotes the qubit to $\ket{e}$, thereby displacing the oscillator potential. The pointer then relaxes towards $x=-x_{0}$, which brings the qubit--pointer transition close to resonance with the battery and enables excitation transfer through $\hat{\sigma}\hat{J}_{+}$. Once the qubit returns to its ground state, the oscillator potential shifts back towards $x=0$, and the reverse battery-to-qubit exchange becomes off-resonant. The pointer thus converts information about the qubit state into a state-dependent modulation of the excitation-transfer rate, reproducing the operational measurement-and-feedback structure of a Maxwell demon without externally performed measurements or time-dependent control.

Finally, we specify the dissipative dynamics. The hot bath drives thermal transitions of the qubit, while the cold bath damps the pointer. For a simple and numerically efficient description, we first describe these processes using a local Markovian master equation,
\begin{equation}
    \dot{\hat{\rho}}
    = -i[\hat{H}, \hat{\rho}]
    + \mathcal{L}_h\hat{\rho}
    + \mathcal{L}_c\hat{\rho},
    \label{eq:master}
\end{equation}
with $\hat{H} = \hat{H}_{\rm SO}+\hat{H}_{\rm SB}+\hat{H}_{\rm B}$.
The two thermal dissipators are
\begin{align}
    \mathcal{L}_h\hat{\rho}
    &= \gamma_h (n_h+1)\,\mathcal{D}[\hat{\sigma}] \hat{\rho}
    + \gamma_h n_h\, \mathcal{D}[\hat{\sigma}^\dagger]\hat{\rho},
    \label{eq:Lh} \\
    \mathcal{L}_c\hat\rho
    &= \gamma_c (n_c+1)\,\mathcal{D}[\hat{b} ] \hat{\rho}
    + \gamma_c n_c\,\mathcal{D}[\hat{b}^ \dagger]\hat{\rho}, \label{eq:Lc}
\end{align}
where $n_h = (e^{\omega_0/T_h}-1)^{-1}$ and $n_c = (e^{\nu/T_c}-1)^{-1}$ are the thermal occupation numbers of the bosonic reservoirs evaluated at the corresponding bare subsystem frequencies. The Lindblad dissipator is defined as
\begin{equation}
\mathcal{D}[\hat A]\hat{\rho}= \hat{A}\hat{\rho}\hat{A}^\dagger- \frac{1}{2}\{\hat{A}^\dagger\hat{A},\hat{\rho}\}.
\label{disi}
\end{equation}

The dissipators in Eqs.~\eqref{eq:Lh} and~\eqref{eq:Lc} describe the thermal contacts of a separate qubit and a separate oscillator degree of freedom, respectively.
 Their use is therefore strictly justified when the internal couplings do not significantly modify the subsystem spectra, approaching the decoupled limit $\lambda,g\to0$ in our setting. At finite coupling strengths $g$ and $\lambda$, the local dissipators provide a numerically efficient approximation that makes simulations with higher-dimensional batteries tractable, but they may lead to quantitative deviations or thermodynamic inconsistencies~\cite{LevyKosloff2014,Hofer2017,cattaneo_local_2019,De_Chiara_2018}.
For the two-level battery considered first in Sec.~\ref{sec:performance}, we therefore evaluate the steady-state performance using a global Born--Markov--secular (BMS) master equation constructed in the eigenbasis of the full interacting Hamiltonian. We subsequently employ the numerically more efficient local description for higher-dimensional batteries and dynamical calculations.
In Sec.~\ref{glob_vs_loc}, we benchmark the local description more systematic treatments; in particular, the textbook global Born--Markov--secular (BMS) master equation, which induces transitions between the energy eigenstates of the combined qubit--oscillator--battery system~\cite{BreuerPetruccione2002}. Our comparison shows that the local dissipators reproduce the relevant steady-state energetics fairly well within the operating regime of our model.

\section{Steady-state performance}\label{sec:performance}

\begin{figure*}
    \centering
     \includegraphics[width=1\linewidth]{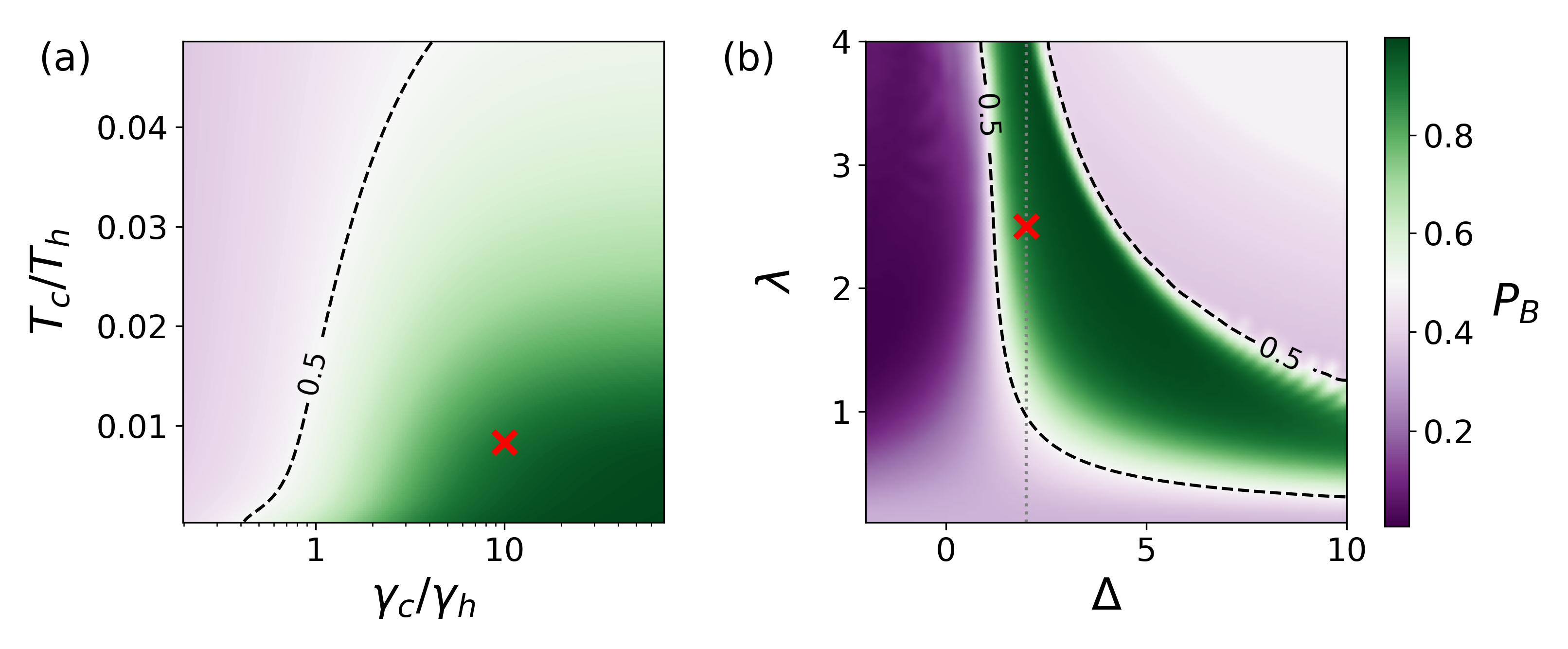}
    \caption{Steady-state excitation probability $P_{\rm B}$ of a two-level battery, calculated using the global Born-Markov-Secular master equation. Population inversion $P_{\rm B} > 1/2$ is indicated in green shades and delimited by the dashed line.
    In (a), we vary the cold-bath temperature $T_c$ and the pointer damping rate $\gamma_c$ for a fixed $\Delta=2$. In (b), we vary the detuning parameter $\Delta$, defined through $\omega_B=\omega_0-\Delta\lambda^2\nu$, and the qubit--pointer coupling strength $\lambda$. The vertical line marks the characteristic resonance $\Delta=2$. Unless varied explicitly, the parameters are $\nu=0.01\omega_0$, $\gamma_h=0.01\nu$, $\gamma_c=0.1\nu$, $g=0.3\nu$, $T_h=\omega_0/\ln2$, $T_c=1.2\nu$, and $\lambda=2.5$. The red cross marks the common reference parameter set across the three panels.}
    \label{fig:occprob_pair_wide}
\end{figure*}

We now assess the performance of the demon engine model through numerical simulations and demonstrate autonomous work extraction in the expected parameter regime. We quantify the work output in this autonomous, i.e., non-cyclic, setting by the achievable steady-state battery charge, defined as the amount of operationally useful energy accumulated in the reduced state of the battery subsystem, $\hat\rho_{\rm B}(t)=\Tr_{\rm S,O}[\hat\rho(t)]$, as $t\to\infty$.
In the absence of an external phase reference, the steady-state battery state has vanishing coherences between the energy eigenstates of $\hat H_{\rm{B}}$, consistent with the phase symmetry of the model (see. Eq.~\eqref{eq:master} and confirmed by our numerical simulations. In all of the following, we fix the exemplary model parameters $\nu=0.01\omega_0$, $\gamma_h=0.01\nu$, and $T_h=\omega_0/\ln2$, such that $n_h=1$.

\begin{figure}
    \centering
    \includegraphics[width=1 \linewidth]{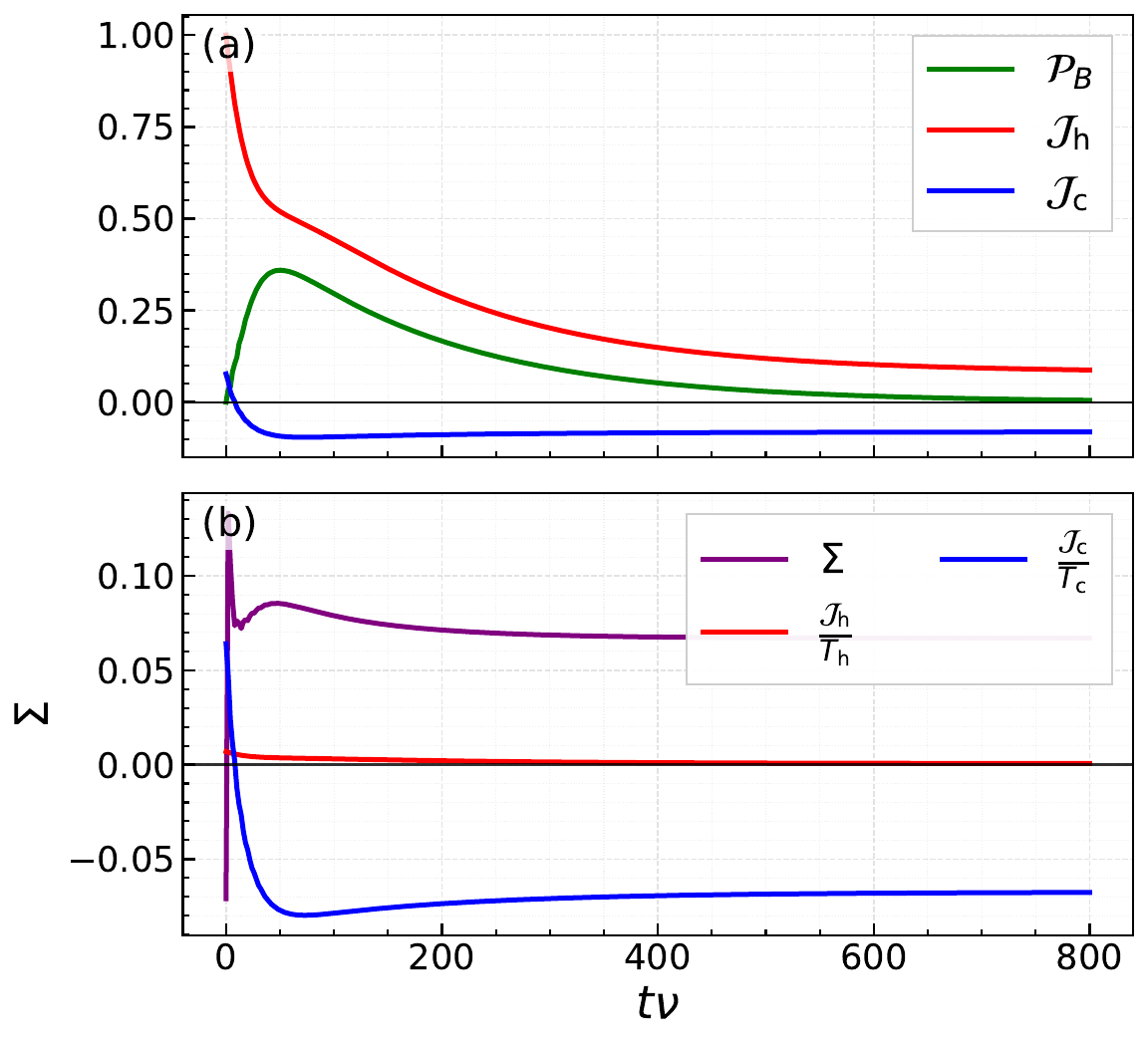}
    \caption{Energy current and entropy production as a function of time. The system evolves according to the local master equation~\eqref{eq:master}, and it is initialized in the ground state of the full Hamiltonian. The model parameters correspond to the red cross in Fig.~\ref{fig:occprob_pair_wide}. (a) Heat currents from the hot (red) and the cold (blue) bath, together with the average battery charging power (green). The dashed lines indicates the steady-state heat currents. (b) Entropy production rate (purple) and the corresponding hot- and cold-bath contribution (red and blue), entering Eq.~\eqref{eq:second_law}.}
    \label{fig:current}
\end{figure}

Let us first consider the instructive case of a two-level battery, $j=1/2$, with energy eigenstates $\ket{m=\pm 1/2}$. Its energy content is determined by the excited-state population, $P_{\rm{B}} = \bra{1/2}\hat\rho_{\rm B}\ket{1/2}\label{eq:expect}$. For an energy-diagonal two-level battery, useful battery charge is indicated by population inversion~\cite{RevModPhys.96.031001}, $P_{\rm{B}} > 1/2$, which corresponds to a state with nonzero ergotropy and may serve, e.g., as a resource for stimulated emission into an appropriate field mode. In Fig.~\ref{fig:occprob_pair_wide}, we show density plots of the steady-state value of $P_{\rm{B}}$ as a function of different model parameters.
The population-inverted operating regime is delimited by the dashed contour, with darker shades of green indicating stronger population inversion.
The steady-state results shown in Fig.~\ref{fig:current} are obtained using the global Born--Markov--secular (BMS) master equation discussed in Sec.~\ref{glob_vs_loc}.
The local approach introduced in Sec.~\ref{sec:model} yields similar results within the relevant operating regime, but can produce small unphysical regions outside it, e.g., population inversion for a blue-detuned battery at very strong qubit-pointer coupling $\lambda$. Since both approaches agree well within the operating regime, we later resort to the numerically more efficient local description for the larger multilevel batteries.

Figure~\ref{fig:occprob_pair_wide}(a) shows the battery excitation as a function of the pointer damping rate $\gamma_{c}$ and the cold bath temperature $T_{c}$. The population-inverted working regime appears when the pointer relaxes sufficently fast compared to the thermally induced qubit transitions, namely for $\gamma_{c}>\gamma_{h}$. This behaviour reflects the basic operating mechanism of the engine. After the qubit is excited by the hot bath, the pointer must relax towards the displaced equilibrium position associated with the excited qubit state. Only then is the qubit-battery transition resonantly selected and energy can be transferred efficiently to the battery. If $\gamma_{c}$ is too small, the pointer does not reliably encode the qubit state before the next qubit transition occurs, and the selective charging mechanism is suppressed.

Figure~\ref{fig:occprob_pair_wide}(b) shows the battery excitation as a function of battery-qubit detuning, parametrized by $\Delta = (\omega_0-\omega_{\rm B})/\lambda^2 \nu$, and the qubit-pointer coupling strength $\lambda$.
 The resonance condition estimated in the previous section, $\omega_{\rm B}\simeq\omega_0-2\lambda^2\nu$, corresponds to $\Delta\simeq2$ and is marked by the vertical line.
  For sufficiently strong $\lambda$, the optimal working regime is indeed found around the expected detuning, and extends towards somewhat larger values of $\Delta$, supporting the proposed state-selective charging mechanism. Conversely, we observe a relative cooling effect on the battery when it is blue-detuned by a comparable amount, $\Delta \lesssim -1$.

  For  weaker qubit-pointer coupling, $\lambda \lesssim 1$, the optimal detuning shifts towards larger values of $\Delta$. We attribute this behavior to the increasing overlap between the thermal states of the displaced and the undisplaced pointer corresponding to the qubit states $\ket{e}$ and $\ket{g}$, respectively. Nevertheless, the selective charging mechanism and population-inverted working regime still exist at $\Delta \gg 1$, as the two states can be discerned by the Gaussian tails of the thermal distribution of pointer positions.

In a practical setting, the performance of an autonomous quantum engine may be assessed not only in terms of the amount of useful energy stored in the battery at steady state, but also by the time required to reach that state, i.e., the mean charging power of the battery system. We will not be concerned with optimising this power in the present work, but clearly, the parameter regimes of optimal work output and optimal charging may differ. For example, small values of the qubit-battery coupling $g \ll \nu$ or large detunings $\Delta \gg 1$ can result in high population inversion, but they will also increase the time to reach this inversion.
We illustrate the charging dynamics in Fig.~\ref{fig:current}(a), where the relevant energy currents are shown as a function of time using the local master equation~\eqref{eq:master}. For simplicity, the system is initialized in the ground state of the full Hamiltonian $\hat{H}$.
The heat currents from the hot and the cold bath, shown in red and blue, respectively are defined as $\mathcal{J}_j(t) = \Tr [\hat{H} \mathcal{L}_j \hat\rho(t)] $ with $j=h,c$.
The green curve shows the instantaneous battery charging power, which we quantify by the rate of change of the mean battery energy, $\mathcal{P}_B(t)=\Tr [\hat H_B\,\dot{\hat \rho}(t)]$.
Following the initial transient, the charging power reaches a maximum around $\nu t\simeq50$ and subsequently decreases as the battery approaches its steady state. At long times, the charging power vanishes, while a finite heat current from the hot to the cold bath remains, corresponding to an idle heat flow once the finite-capacity battery is charged.

In Fig.~\ref{fig:current}(b), we assess the thermodynamic consistency of the local description in terms of entropy production rate,
\begin{equation}
    \Sigma(t) = \dot{S}(t) - \frac{\mathcal{J}_{\mathrm{h}}(t)}{T_{\mathrm{h}}}
    - \frac{\mathcal{J}_{\mathrm{c}}(t)}{T_{\mathrm{c}}},
    \label{eq:second_law}
\end{equation}
where $\dot{S} (t) = -\Tr[ \dot{\hat \rho} \ln \hat{\rho}]$ is the rate of change of von Neumann entropy.
For a thermodynamically consistent master equation in which each reservoir separately thermalizes the system to the corresponding Gibbs state, the entropy production rate is non-negative, $\Sigma(t)\geq0$~\cite{Alicki_1979,Spohn1978}.
The global BMS master equation satisfies this property, whereas is not guaranteed in general for the approximate local dissipators~\eqref{eq:Lh} and~\eqref{eq:Lc}.

Nevertheless, for the representative parameter set shown in Fig.~\ref{fig:current}, the local description yields positive entropy production after the initial transient and, in particular, at steady state. We find positive steady-state entropy production throughout the relevant operating regime, providing an additional consistency check for the use of the local master equation~\eqref{eq:master} in this parameter range.

Let us now consider larger multilevel batteries and ask how the amount of useful stored energy depends on the battery size $n=2j+1$.
At steady state, the energy-diagonal reduced battery state is characterised by the populations $p_m = \bra{m} \hat\rho_{\rm B}\ket{m}$ associated to the energy eigenvalues $\epsilon_m = \omega_{\rm B} (m+j)$.
The useful energy contained in this population distribution is conveniently quantified in terms of the ergotropy~\cite{Allahverdyan2004,Choquehuanca2025}. It is defined as the maximum energy that can be extracted on average by applying a unitary operation, here to the reduced state of the battery,
\begin{equation}
\mathcal{W} = \max_{\hat V \text{ unitary}} \left\{ \text{tr}[\hat H_{\rm B}\hat{\rho}_{\rm B}] - \text{tr}[\hat H_{\rm B} \hat V \hat{\rho}_{\rm B} \hat V^\dagger] \right\} .
\end{equation}
Thermodynamically speaking, the ergotropy represents the maximum work output in an isentropic process with no net change of the battery Hamiltonian $\hat H_{\rm B}$.
For the battery Hamiltonian~\eqref{eq:HB}, the ergotropy can take values between zero and the maximum battery energy, $ 0\le \mathcal{W} \le \omega_{\rm B}$.

Given the ascending sequence $(\epsilon_m)_m$ of battery energies, $\epsilon_{m+1} \geq \epsilon_m$ $\forall m$, and the re-ordered sequence $(p_m^{\downarrow})_m$ of populations in descending order, $p_{m+1}^{\downarrow} \leq p_m^{\downarrow}$ $\forall m$, the ergotropy for an energy-diagonal battery state $\hat\rho_{\rm B}$ can be given explicitly as the difference between the mean energy contents of this state and the so-called passive state $\hat\pi_{\rm B} = \sum_m p_m^{\downarrow} \ket{m}\!\bra{m} $,
\begin{equation}
\mathcal{W} = \text{tr}[\hat H_{\rm B}\hat{\rho}_{\rm B}] - \text{tr}[\hat H_{\rm B}\hat{\pi}_{\rm B}] = \sum_{m=-j}^j \epsilon_m (p_m - p_m^{\downarrow}). \label{eq:ErgoDiag}
\end{equation}
The energy remaining in the passive state cannot be extracted by a unitary operation on the battery alone. In particular, any energy-diagonal state with populations that decrease monotonically with increasing energy, including a Gibbs state, has zero ergotropy. For the two-level battery considered above, $\mathcal{W} = \omega_{\rm B} \max \{2P_{\rm B}-1,0 \}$ which makes explicit the connection between population inversion and useful battery charge in the two-level case.

Here, we evaluate the ergotropy of the reduced battery state only, since we regard the energy stored locally in the battery as the accessible output of the engine. The ergotropy of the full system may additionally contain extractable energy stored in the working medium, the pointer, their correlations, or the interaction energy, which we do not count as battery charge.
 In our numerical results, however, the importance of these additional contributions relative to the ergotropy stored in the battery decreases as the battery size increases.

\begin{figure}
    \centering
    \includegraphics[width=1\linewidth]{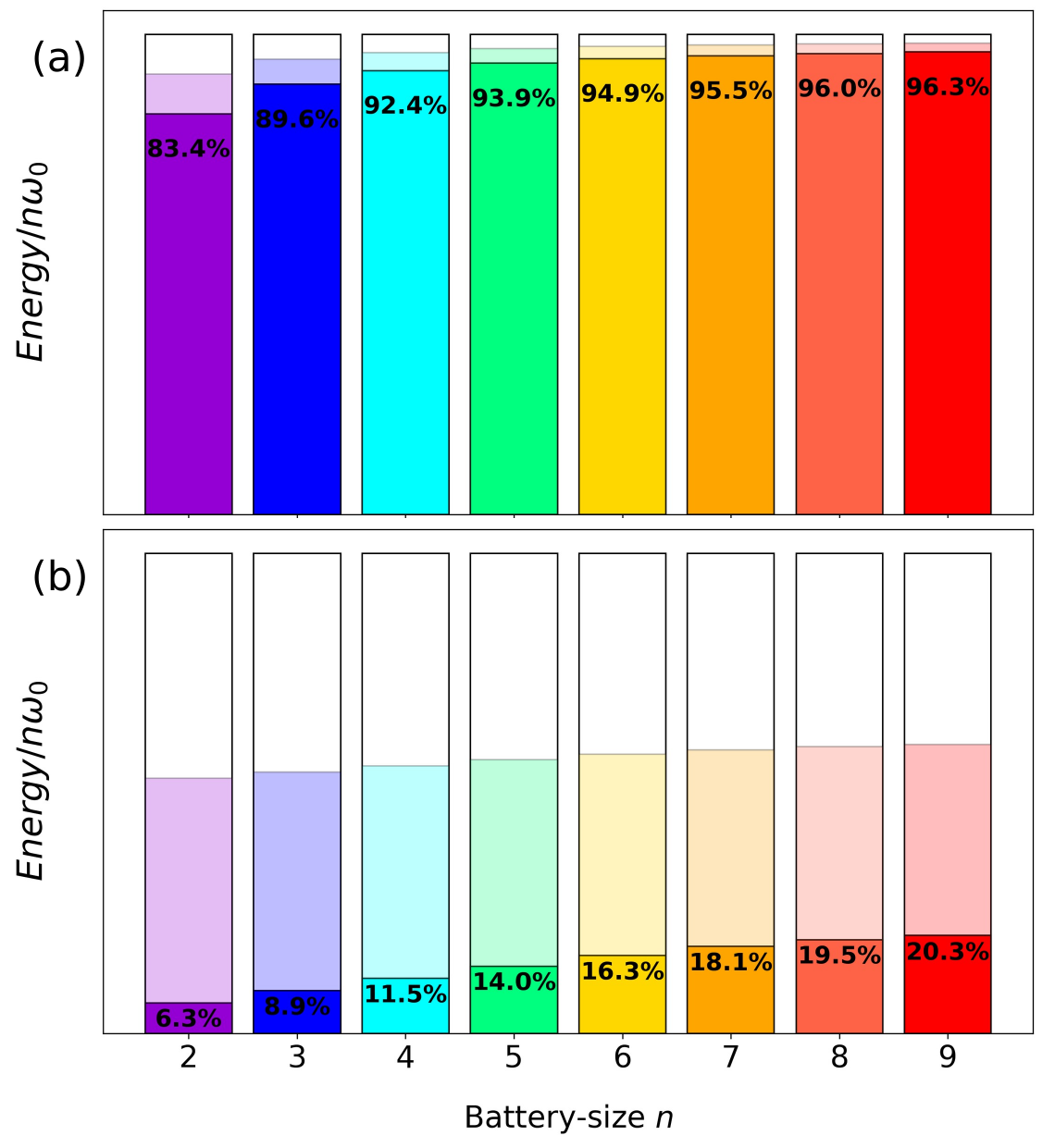}
    \caption{Steady-state charge level as a function of the number $n$ of battery levels, for (a) the common parameter settings of Fig.~\ref{fig:occprob_pair_wide} and (b) a five-fold increased cold bath temperature, $T_c = 6\nu$. The filled bars show the ergotropy $\mathcal W$ relative to the maximum battery energy $(n-1)\omega_{\rm B}$, while the semitransparent bars show the correspondingly normalized mean battery energy.}
    \label{Battery}
\end{figure}

Figure~\ref{Battery} compares the steady-state charge for numerically accessible battery sizes from $n=2$ to $9$. We characterize the relative useful charge by the ergotropy normalized to the maximum battery energy, $0 \leq \mathcal{W}/n\omega_B \leq 100\%$. Panel (a) corresponds to the reference parameter regime of Fig.~\ref{fig:occprob_pair_wide}, whereas in (b) we consider a five-fold larger cold-bath temperature, $T_c = 6 \nu $ which results, as expected, in a lower relative charge of the battery. The semitransparent bars show the mean energy content of the battery normalized by the maximum energy $n\omega_{B}$ achievable by the battery. This quantity is barely above the ergotropy charge level in (a), but dominant in (b). At the lower cold-bath temperature, most of the energy stored in the battery is ergotropic and therefore available for work extraction. At the higher temperature, the mean battery energy remains appreciable, while a larger fraction of it is passive, resulting in a lower relative ergotropy.
Overall, the absolute amount of useful energy stored in the battery increases with its size, while the relative charge tends towards saturation for larger $n$.

\begin{figure}
    \centering
    \includegraphics[width=1\linewidth]{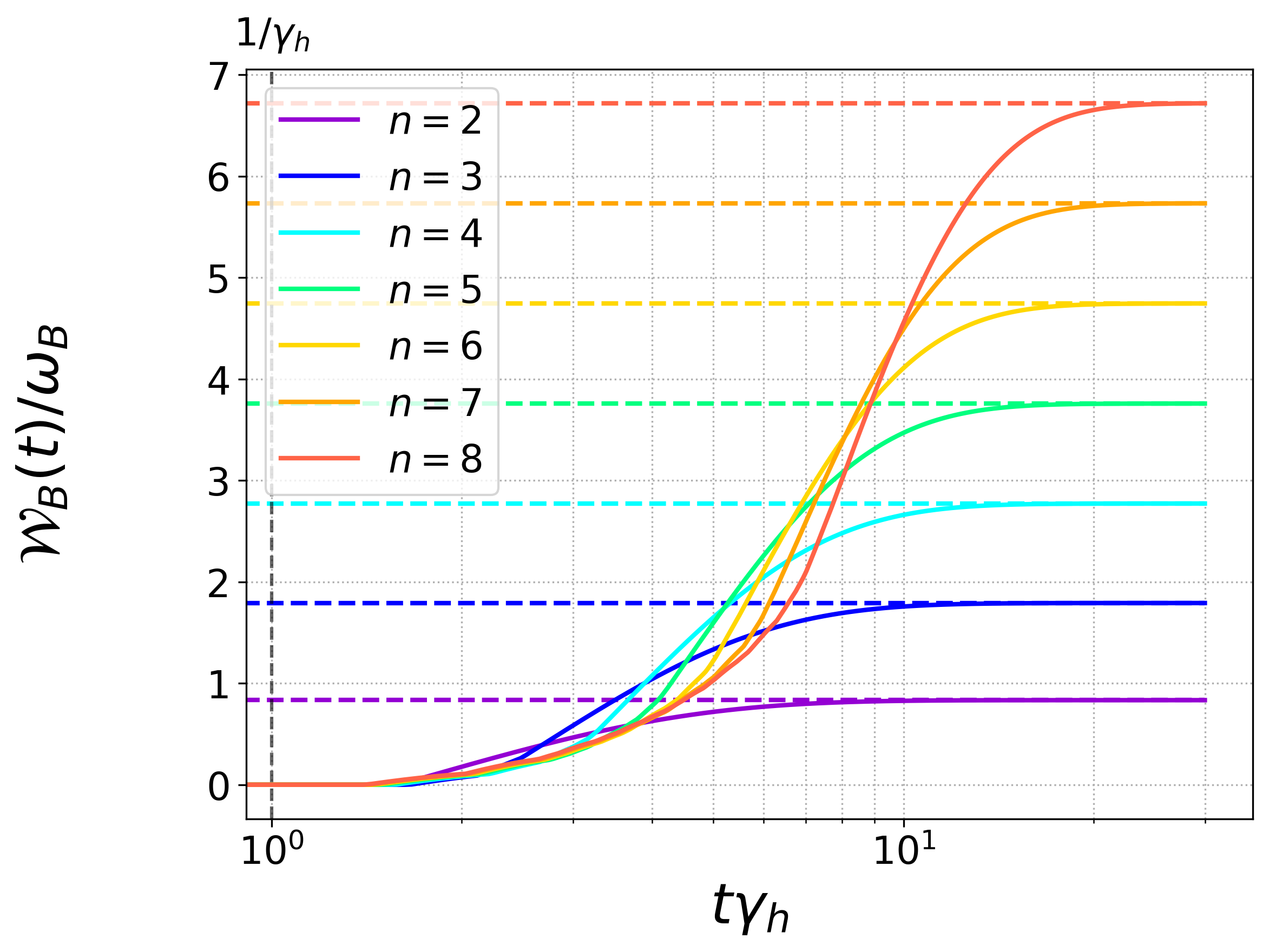}
    \caption{Battery ergotropy as a function of time for multilevel batteries of various sizes $n$. The asymptotic steady-state values are marked by the dashed lines. The model parameters are those of Fig.~\ref{Battery}(a), and the combined system is initialized in the ground state of the full Hamiltonian.
    }
    \label{fig:Ergo-Hfree}
\end{figure}

Figure~\ref{fig:Ergo-Hfree} shows the charging dynamics (on logarithmic scale) for batteries of different sizes $n$.
Larger batteries can accumulate more ergotropy, but require progressively longer times to approach their steady-state charge.  The battery begins to charge only after the qubit has been thermally excited by the hot bath, which for the parameters considered here occurs on a characteristic timescale of order $1/\gamma_h$. For larger batteries, repeated excitation-transfer events are required to populate higher battery levels, and several such timescales are therefore needed before a significant amount of ergotropy can accumulate.

In all numerical simulations, the oscillator Hilbert space was truncated at excitation number $39$, resulting in a total Hilbert-space dimension of $80 n$ for the full model. We represented the density operator $\hat\rho$ as a column vector and the generators of both the local master equation~\eqref{eq:master} and the global BMS equation as sparse matrices. Time evolutions were computed by evaluating the action of the corresponding matrix exponential on the vectorized state using the \texttt{expm\_multiply}  routine from the  \texttt{scipy.sparse.linalg}  package, thereby avoiding explicit construction of the full exponential. Steady states were obtained from the null space of the corresponding generator. For the global BMS description, we further exploited the fact that, for the non-degenerate spectra considered here, the steady state is diagonal in the energy eigenbasis, allowing the steady-state calculation to be restricted to the level populations.

\section{Master equation models}\label{glob_vs_loc}

Most of the quantitative results presented above were obtained using the local master equation~\eqref{eq:master}, whose dissipators describe separate thermal contacts of the qubit with the hot bath and of the oscillator with the cold bath~\cite{BreuerPetruccione2002}.
 As discussed in Sec.~\ref{sec:model}, this description neglects the modification of the subsystem eigenstates and transition frequencies induced by the internal couplings $g$ and $\lambda$, and therefore becomes approximate when these interactions become non-negligible.
 Here we asses the accuracy of the local model by comparing it with three alternative open-system descriptions: the global Born--Markov--secular (BMS) master equation, a hybrid model that accounts for the qubit--pointer interaction in the cold-bath dissipator, and a second-order time-convolutionless (TCL2) master equation.

In all approaches, the reservoirs are modelled as continuous spectra of harmonic oscillator modes at thermal equilibrium, and are linearly coupled to the ststem.
Each bath $j\in{h,c}$ is characterized by a temperature $T_j$ that sets the thermal occupation numbers for every spectral component $\omega$ as $n_j (\omega) = [\exp(\omega/T_j)-1]^{-1}$ and by a spectral density of system-bath coupling rates $J_j(\omega)$. We choose Ohmic spectral densities with an exponential cutoff at ten times  the bare subsystem resonance addressed by each bath,
\begin{equation}
    J_h (\omega) = \eta_h \omega e^{-\omega/20\omega_0}, \quad J_c (\omega) = \eta_c \omega e^{-\omega/20\nu}.
\end{equation}
The prefactors $\eta_h$ and $\eta_c$ are chosen such that the spectral densities reproduce the rate parameters of the local master equation at the corresponding bare subsystem frequencies, $J_h(\omega_0) = \gamma_h$ and $J_c(\nu) = \gamma_c$.

The global BMS master equation is constructed in the eigenbasis of the full system Hamiltonian $\hat H$ according to the standard weak-coupling Born--Markov--secular treatment~\cite{BreuerPetruccione2002}. Each bath acting individually induces thermodynamically consistent incoherent transitions between the system energy eigenstates and the decay of energy coherences, driving the system towards the Gibbs state at the corresponding bath temperature. In the simultaneous presence of both baths, the system instead approaches a nonequilibrium steady state that can produce finite battery ergotropy, while the steady-state entropy production~\eqref{eq:second_law} remains consistent with the second law~\cite{Alicki_1979}.

We expect the global BMS description to be more reliable than the local model when the internal interactions significantly modify the system eigenstates. In the present setting, the dominant effect arises from the strong qubit--pointer coupling, $\lambda\gtrsim1$.
Since the qubit-battery coupling remains fairly weak one may consider an intermediate hybrid description in which the qubit--pointer interaction is incorporated into the cold-bath dissipator, while the hot bath and the battery coupling are treated as in the local model.

In this hybrid description, the cold bath thermalizes the oscillator towards a qubit-dependent equilibrium state: a Gibbs state that is conditionally displaced by $-x_0$ when the qubit is excited. The resulting dissipator is obtained by replacing the mode operator $\hat b$ and its conjugate in the cold bath dissipator~\eqref{eq:Lc} with $\hat b + \lambda \hat\sigma^\dagger \hat\sigma$ and its conjugate.
At the same time, the hot bath is treated locally as we assume $\omega_0 \gg \nu$ and omit the oscillator-induced energy modulation.
Such hybrid descriptions have been employed for vibronic molecular systems \cite{Alicki2019MolecularBattery,Zhang2024} as well as for the non-autonomous demon engine scheme in Ref.~\cite{Seah2020}. However, since the hybrid model omits the battery subsystem alltogether, it may not necessarily lead to improved results in the present setting.

Finally, we consider as the most accurate reference model the second-order time-convolutionless (TCL2) master equation, which represents the leading-order perturbative expansion of the system state dynamics in the weak system-bath coupling limit. It supersedes the global BMS model, because it does not invoke the Markov approximation and retains finite-time bath-correlation integrals, leading in general to a dynamics that is not necessarily of Lindblad form. The TCL2 equation is commonly stated in the interaction picture with respect to the full system Hamiltonian $\hat H$ in which system operators $\hat A$ including the quantum state are represented by $\hat{A}_I(t) = e^{i\hat H t}\hat A e^{-i\hat H t}$ \cite{BreuerPetruccione2002}. Here,
\begin{align}
    \frac{d}{dt}\hat\rho_I(t) ={}& \sum_{j=h,c}\int_0^t d\tau \, C_j (\tau) \left[ \hat{A}^{(j)}_I (t-\tau) \hat\rho_I(t) \hat{A}^{(j)}_I (t) \right. \nonumber \\
    &\left.- \hat{A}^{(j)}_I (t) \hat{A}^{(j)}_I (t-\tau) \hat\rho_I (t) \right] + \text{h.c.},
\end{align}
with $\hat{A}^{(h)} = \hat{\sigma}_x$ and $\hat{A}^{(c)} = \hat{b} + \hat{b}^\dagger$. The respective bath correlation functions are given as
\begin{equation}
    C_{j}^{}(\tau)=\int_0^{\infty}\!\! d \omega \frac{J_j(\omega)}{\pi}\left( [n_{j}(\omega)+1]e^{-i\omega\tau} +n_{j}(\omega) e^{i \omega \tau} \right).
\end{equation}

\begin{figure}
    \centering
    %d\hspace*{-1.5cm}
    \includegraphics[width=\columnwidth]{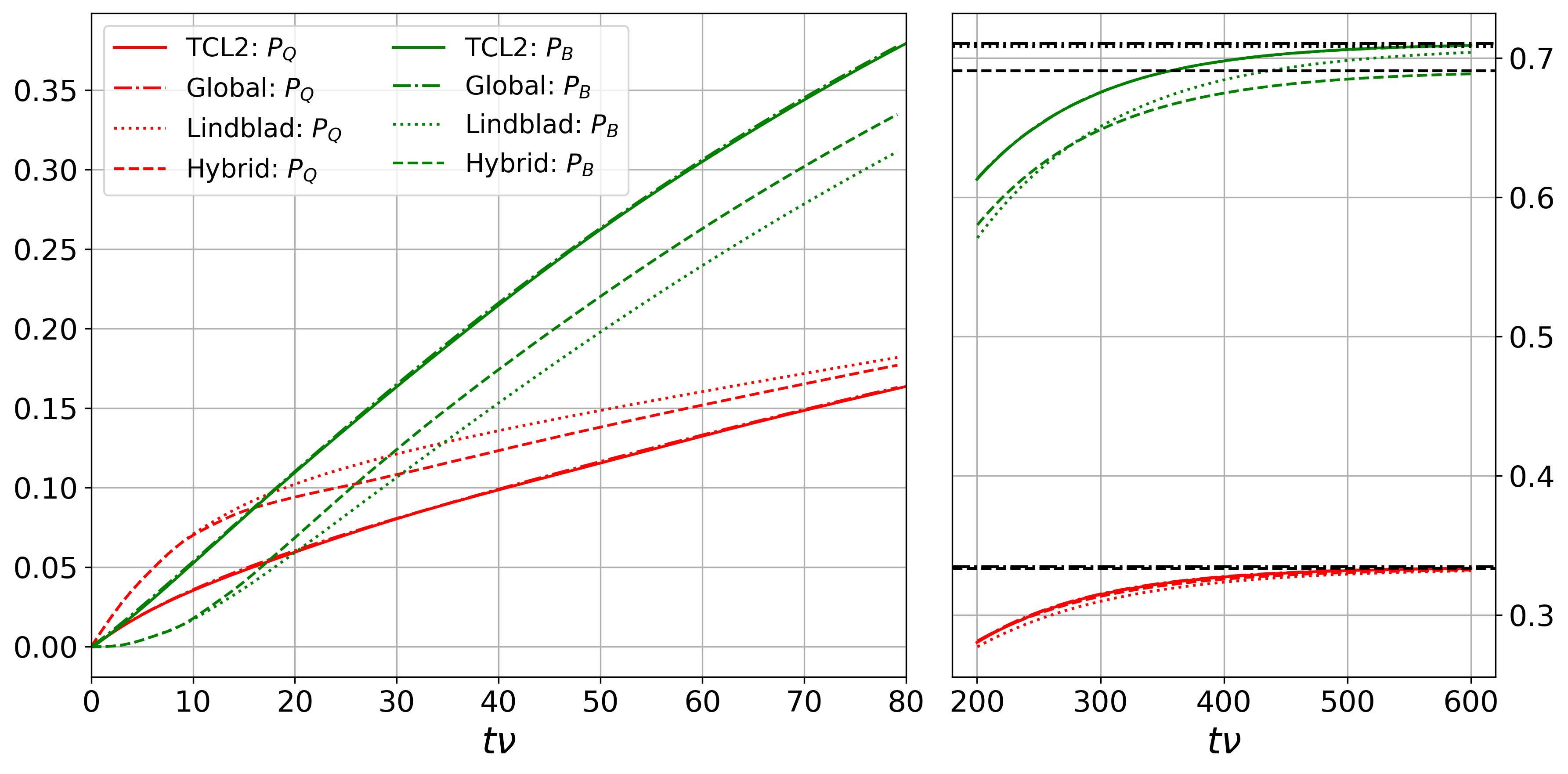   }
    \caption{Time evolution of the excited-state population of the battery $P_{\rm B}$ and of the qubit $P_{\rm Q}$, obtained from th four master euqations described in the main text.
    The solid line shows the TCL2 dynamics, the dash-dotted line
    the global Born--Markov--secular dynamics, the dashed line the hybrid model,
    and the dotted line the local master equation~\eqref{eq:master}.
    The decay rates, as well as $\nu$ and $\omega_0$, are chosen as in
    Fig.~\ref{fig:occprob_pair_wide}. With battery frequency $\Omega_B = \omega_0 - 3\lambda^2 \nu$. The combined system is initialized in the ground state of the full Hamiltonian.}
    \label{red_vs_lind}
\end{figure}

To assess the accuracy of the local, the hybrid, and the global BMS master-equation models, we compare their predictions with the TCL2 dynamics. Because the TCL2 and global BMS calculations are numerically demanding, we restrict this comparison to a two-level battery and a representative parameter set within the operating regime. We consider the excited-state populations of the battery and qubit, $P_{\rm B} (t)$ and $P_{\rm Q} (t) = \Tr[\hat\sigma^\dagger\hat\sigma \hat\rho(t)]$.
The simulation results are shown in Fig.~\ref{red_vs_lind} with panel (a) focusing on short times and panel (b) on the approach of the steady state.
The global BMS dynamics closely follows the TCL2 result at all times, and the two respective curves are barely distinguishable. At short times, the local and hybrid models show somewhat larger deviations. They remain close to one another, indicating that incorporating the qubit-dependent displacement only in the cold-bath dissipator does not substantially modify the dynamics for the parameter set considered here.
At longer times, the local and hybrid descriptions approach similar stationary populations as BMS and TCL2. The agreement is particularly close for the qubit population, while somewhat larger deviations remain for the battery population. Nevertheless, all four descriptions predict the same qualitative operating behaviour and a population-inverted battery in the parameter regime considered.

\section{Possible platforms}\label{sec:platforms}

In this section, we discuss possible experimental platforms on which to realize the autonomous demon engine. Suitable platforms must have the following essential ingredients: an effective two-level system and an oscillator degree of freedom that are coupled through a Holstein-type Hamiltonian sufficiently strongly ($\lambda \gtrsim 1$), a hot source of thermal excitations of the two-level system that are slow compared to the relaxation of the oscillator by a cold environment ($\gamma_c > \gamma_h$), and a long-lived auxiliary spin system that can weakly exchange excitations with the two-level system at a controllable detuning ($\omega_{\rm B} \simeq \omega_0 - 2\lambda^2 \nu$). In the following, we focus on two representative physical systems with which one might be able to satisfy these criteria.

One possible experimental implementation could involve vibronic molecules, in which the coupling between the electronic state and the vibrational modes of the nuclei leads to a Franck-Condon shift~\cite{Franck1924,Clark1986, Zhang2021} between absorption and emission peaks in fluorescence spectroscopy. This shift is related to a change in the equilibrium positions of the nuclei upon excitation, which results in a net displacement of the associated nuclear vibrations as described by the Holstein Hamiltonian; see Appendix \ref{app:HolsteinMolecules}. A pronounced shift corresponds to a strong displacement and coupling parameter $\lambda$, as required for our purposes.
Drawing on the findings in Ref.~\cite{Frank1998}, we identify formaldehyde as a promising example. It features a strong coupling between the dominant electronic transition ($\omega_0 \approx 2\pi\times 850\,$THz) and the vibration mode associated to the $\ce{C=O}$ bond that results in a measured increase of the equilibrium bond length by about $12\,$pm. Given the vibrational eigenfrequency $\nu \approx 2\pi \times 52\,\mathrm{THz}$~\cite{AlDerzi2021improved} and the reduced oscillator mass $\mu \approx 6.9\,\mathrm{u}$, which amounts to a ground-state oscillation amplitude of about $5.3\,$pm, we estimate a suitable coupling parameter $\lambda \approx 1.6$.

The main obstacle towards realizing the autonomous demon engine with vibronic molecules such as formaldehyde is the fact that the lifetimes of vibrational excitations are usually much longer than electronic lifetimes, i.e., $\gamma_c \ll \gamma_h$. The electronic linewidth for isolated formaldehyde molecules was reported as $\gamma_h \approx 6.2\,$MHz~\cite{Weisshaar1979}, whereas the vibrational relaxation rate was calculated to be as low as $\gamma_c \approx  30\,$Hz~\cite{AlDerzi2021improved}. A possible route towards bridging this huge gap is Purcell enhancement of the vibrational decay rate, which has been demonstrated up to a factor of $10^6$ with help of infrared nano-antennas~\cite{PhysRevLett.123.153001}. For our purposes, one additionally needs a spectrally filtered heat source and a long-lived multilevel battery system exchanging excitations with the electronic state. To this end, one could embed the molecule in a high-finesse optical cavity, as suggested in \cite{PhysRevLett.122.203602}.

A second promising platform is the electromechanical setup proposed in Ref.~\cite{martinetz_quantum_2020}, in which a harmonically trapped, charged nanoparticle couples capacitively to a Cooper-pair box. Here, the former acts as the pointer while the latter plays the role of the qubit. An additional Josephson junction device or high-finesse microwave resonator with a tunable resonance could serve as the battery and extract excitations from the qubit. Typical Josephson qubit transition frequencies $\omega_0$ are of the order of GHz, and their linewidths $\gamma_h$ can reach several kHz or even less in state-of-the-art implementations~\cite{Burnett_2019,tuokkola_methods_2025}. The trapping frequency $\nu$ is controlled by the applied trap voltage and the charge-to-mass ratio of the nanoparticle. The proposed setup parameters amount to $\nu \approx 138\,$kHz and facilitate ultrastrong qubit-oscillator coupling strengths of up to $\lambda \approx 122$. At the same time, the large separation between $\omega_0$ and $\nu$ ensures that the shifted resonance $\omega_B \simeq \omega_0 - 2\lambda^2\nu$ remains spectroscopically resolvable.

Once again, the requirement $\gamma_c>\gamma_h$ is not satisfied here, as the proposed oscillator damping rate is $\gamma_c \approx 163\,$Hz. A heavier (and more highly charged) nanoparticle than the proposed mass of $10^6\,$u would grant access to greater damping rates with help of state-of-the-art optical cooling techniques for nanoparticles such as feedback cooling \cite{Tebbenjohanns2021}.

\section{Conclusions \& Outlook}\label{sec:conclusion}
We have introduced an autonomous demon engine model based on a Holstein-type coupling between a harmoinc oscillator acting as a pointer and a qubit that receives thermal excitations from a hot thermal bath. The Holstein coupling induces a qubit state-dependent displacement of the pointer equilibrium position, which in turn generates an effective frequency shift of the qubit. This state-dependent shift enables selective energy transfer from the hot bath into a multi-level spin battery system. The pointer effectively performs both the measurement and the feedback step of a conventional demon engine, converting heat into useful battery charge, such that no external measurements or control operations are required. Neither does the model require an explicit erasure of the ``demon's memory'' as thermodynamic consistency is established by a cold thermal bath damping the pointer. Thus the demon is exorcised.

Our numerical analysis of the model predicts a stationary battery charge in the form of population inversion, as quantified by ergotropy, in a parameter regime of strong qubit-oscillator coupling, weak qubit-battery coupling, and fast oscillator damping compared to the thermal excitation rate of the qubit.
We corroborate the validity of our results by finding a good agreement between the predictions of  different dynamical descriptions of our model: one based on the phenomenological local master equation, one based on the global Born-Markov-secular master equation, and one based on the leading-order time-convolutionless master equation. Hence the proposed work production mechanism remains robust beyond idealized assumptions.
We also explored the possibility of physical implementations with vibronic molecules such as formaldehyde and in a proposed electromechanical setup with a trapped nanoparticle coupled to a superconducting qubit. Both platforms are promising, but require additional methods to enhance the effective oscillator damping rate.

Future research could be directed towards concrete realizations of the demon engine scheme with vibronic molecules, optomechanical systems, or other on other suitable experimental platforms. To this end, one could employ an optical cavity in place of the battery system so as to directly generate an electromagnetic field as the useful output of the engine \cite{Monsel2018}.
On the theory side, it would be valuable to investigate non-Markovian effects at strong bath coupling using numerically exact methods such as hierarchical equations of motion \cite{w3nw-hbjc} or higher-order time-convolutionless master equation expansions. Finally, the coupling of several demon units opens opportunities for studying collective phenomena, synchronization, and scalable autonomous heat machines.

%da noch bbl file dazu 

\onecolumngrid
\newpage
\appendix
\section{Franck-Condon sidebands and the battery resonance condition}
\label{app:FC_sidebands}

In the main text, we argue that the battery frequency should be detuned from the qubit resonance by
resonance $\omega_B \simeq \omega_0-2\lambda^2\nu$ in order to achieve work extraction.
Here we derive this condition from the underlying Holstein Hamiltonian of the qubit-pointer subsystem $\hat H_{\rm SO}$, as given in \eqref{eq:HSO}. We use $\hbar=1$, as in the main text.
The first step is to notice that the qubit-pointer Hamiltonian can be rewritten as
\begin{equation}
    \hat{H}_{\mathrm{SO}} =
    (\omega_0-\lambda^2 \nu)\hat{\sigma}^{\dagger}\hat{\sigma}
    + \nu\left(\hat{b} + \lambda \hat\sigma^\dagger\hat\sigma \right)^{\dagger}\left(\hat{b} + \lambda \hat\sigma^\dagger\hat\sigma \right),
    \label{app:eq:HSO}
\end{equation}
which indicates that the excited vibronic manifold is shifted by the energy $-\lambda^2\nu$. The Hamiltonian can be diagonalized by a polaron transformation, described by the unitary
\begin{equation}
\hat{U} = e^{\lambda(\hat{b}^\dagger-\hat{b})\hat{\sigma}^\dagger\hat{\sigma}} = \mathbb{1} \otimes \ket{g}\bra{g} + \hat D(\lambda) \otimes \ket{e}\bra{e},
\end{equation}
with $\hat{D}(\lambda) = \exp [\lambda\hat{b}^{\dagger}- \text{h.c.}]$ the usual displacement operator. Noting that $\hat{D}^\dagger (\lambda) =\hat{D}(-\lambda)$ and $\hat D(\lambda) \hat b \hat D^\dagger (\lambda) = \hat b - \lambda$, we find that the transformed Hamiltonian in the polaron frame is decoupled and therefore diagonal,
\begin{equation}
    \tilde{H}_{\mathrm{SO}}
    =
    \hat{U}\hat{H}_{\mathrm{SO}}\hat{U}^{\dagger}
    =
    (\omega_0-\lambda^2\nu)
    \hat{\sigma}^{\dagger}\hat{\sigma}
    +
    \nu\hat{b}^{\dagger}\hat{b} .
    \label{app:eq:HSO_polaron}
\end{equation}
The associated eigenenergies are
\begin{equation}
    E_{g,m}=m\nu,
    \qquad
    E_{e,n}=\omega_0-\lambda^2\nu+n\nu,
    \label{app:eq:vibronic_energies}
\end{equation}
where $m,n\in\mathbb{N}_0$ label oscillator excitations. In the original frame, the excited-state eigenvectors are the displaced oscillator states $\hat{U}^{\dagger}\ket{e,n}$. A transition between the two vibronic manifolds corresponds to the frequencies
\begin{equation}
    \omega_{n\rightarrow m}
    =
    E_{e,n}-E_{g,m}
    =
    \omega_0-\lambda^2\nu+(n-m)\nu .
    \label{app:eq:sideband_frequency}
\end{equation}
The qubit transition is therefore resolved into a ladder of vibronic sidebands, rather than shifted to a single transition frequency.

The Franck-Condon factors enter through the qubit--battery interaction \eqref{eq:HSB}, which transforms into the polaron frame as

\begin{equation}
    \tilde{H}_{\mathrm{SB}}
    =
    \hat{U}\hat{H}_{\mathrm{SB}}\hat{U}^{\dagger}
    =
    g\left(
        \hat{\sigma}\hat{D}^{\dagger}(\lambda)\hat{J}_{+}
        +
        \hat{\sigma}^{\dagger}\hat{D}(\lambda)\hat{J}_{-}
    \right).
    \label{app:eq:HSB_polaron}
\end{equation}

The appearance of the displacement operator $\hat{D}(\lambda)$ means that a qubit flip is accompanied by a possible change of oscillator state. To see this explicitly, consider the process in which the qubit transfers one excitation to the battery, $\ket{e,n,r} \longrightarrow \ket{g,m,r+1}$, where $r$ labels a battery spin state. By virtue of Fermi's golden rule (for weak coupling strengths $g$), the corresponding transition probability is proportional to the squared matrix element
\begin{equation}
    \left|\bra{g,m,r+1}
    \tilde{H}_{\mathrm{SB}}
    \ket{e,n,r} \right|^2 =
    \left| g\, \bra{m}\hat{D}^{\dagger}(\lambda)\ket{n}
    \bra{r+1}\hat{J}_{+}\ket{r} \right|^2 \propto g^2 F_{mn},
    \label{app:eq:matrix_element}
\end{equation}
where the Franck-Condon factor is given as the oscillator overlap, $F_{mn}= |\bra{m}\hat{D}(\lambda)\ket{n}|^2$.

For the resonance estimate used in the main text, it is sufficient to consider the low-temperature case in which the oscillator is initially close to the ground state of one vibronic manifold, $n=0$. This results in the standard Franck-Condon progression,
\begin{equation}
    F_{m0}
    =
    \left|
        \bra{m}\hat{D}(\lambda)\ket{0}
    \right|^2
    =
    e^{-\lambda^2}\frac{\lambda^{2m}}{m!} .
    \label{app:eq:FC_poisson}
\end{equation}
which is a Poisson distribution in $m$ with mean $\lambda^2$. Therefore, for large $\lambda$, the strongest sideband lies near $m\simeq \lambda^2$.
Combining this result with the vibronic transition frequency in Eq.~\eqref{app:eq:sideband_frequency}, with $n=0$ and $m\simeq\lambda^2$, gives
\begin{equation}
    \omega_{\mathrm{peak}}
    \simeq
    \omega_0-\lambda^2\nu-\lambda^2\nu
    =
    \omega_0-2\lambda^2\nu ,
    \label{app:eq:FC_peak_frequency}
\end{equation}
which is the battery resonance condition stated in the main text. The Franck-Condon progression itself is a standard result for displaced harmonic oscillators. The model-specific step here is to use its dominant sideband as the transition selected by the spin battery.

For a thermally occupied pointer, transitions can start from arbitrary oscillator states $\ket{n}$. The sideband spectrum can then be written as
\begin{equation}
    S_{\mathrm{FC}}(\omega)
    =
    \sum_{n,m=0}^{\infty}
    p_n F_{mn}
    \delta\left[
        \omega-
        \left(\omega_0-\lambda^2\nu+(n-m)\nu\right)
    \right],
    \label{app:eq:FC_spectrum}
\end{equation}
where $p_n$ is the oscillator occupation probability, expressed in the
polaron-frame basis, before the transition. Finite temperature therefore
broadens the sideband response. The closed form of $F_{mn}$ is known
analytically in terms of associated Laguerre polynomials, but for the resonance estimate used here, only the $n=0$ Franck-Condon progression is needed.

\section{Vibronic molecules and Franck-Condon shift}\label{app:HolsteinMolecules}
Here we show how the Holstein Hamiltonian used in the main text arises from the Franck-Condon physics of a molecular system with two electronic states and one dominant vibrational mode the parameters of which are affected by the electronic configuration.
Let $\ket{g}$ and $\ket{e}$ denote the electronic ground and excited states characterized by energies $\omega_g$ and $\omega_e$, and let $\hat{x}$ be a vibrational coordinate with effective mass $\mu$. Depending on the electronic state, the vibrational dynamics is governed by harmonic potentials with distinct frequencies $\nu_g$ and $\nu_e$ as well as distinct equilibrium positions. Assuming that $\hat{x}$ is defined with respect to the equilibrium position in the electronic ground state, we denote by $R_{eg}$ the shifted equilibrium position in the excited state.
%In general, these potentials differ not only in curvature but also in their equilibrium positions. Within this framework,
The total Hamiltonian describing the coupled electron-vibration system then takes the form ($\hbar = 1$),
\begin{align}
\hat H_{\rm mol}
&=
\left[
\omega_e + \frac{\hat{p}^2}{2\mu} + \frac{1}{2} \mu \nu_e^2 (\hat{x} - R_{eg})^2
\right] \hat{\sigma}^\dagger \hat{\sigma}
+
\left(
\omega_g + \frac{\hat{p}^2}{2\mu} + \frac{1}{2} \mu \nu_g^2 \hat{x}^2
\right) \hat{\sigma} \hat{\sigma}^\dagger \nonumber \\
&= \omega_g \hat{\sigma} \hat{\sigma}^\dagger
+ \left( \omega_e + \frac{1}{2}\mu \nu_e^2 R_{eg}^2 \right) \hat{\sigma}^\dagger \hat{\sigma}
+ \frac{\hat{p}^2}{2\mu}
+ \frac{1}{2} \mu \nu_g^2 \hat{x}^2
+ \frac{1}{2} \mu ( \nu_e^2 - \nu_g^2 ) \hat{x}^2 \hat{\sigma}^\dagger \hat{\sigma}
- \mu \nu_e^2 R_{eg} \hat{x} \hat{\sigma}^\dagger \hat{\sigma},
\end{align}
where we have used that $\hat{\sigma}^\dagger \hat{\sigma} + \hat{\sigma} \hat{\sigma}^\dagger = \mathbb{1}_2$, the identity operator on the electron subspace.
Replacing the quadrature operators $\hat{x}$ and $\hat{p}$ with bosonic ladder operates, $\hat{x} = \sqrt{\frac{1}{2 \mu \nu_g}} (\hat{b}+\hat{ b}^\dagger)$ and $
\hat{p} = i \sqrt{\frac{ \mu \nu_g}{2}} (\hat{b}^\dagger - \hat{b})$ with  $[\hat{b},\hat{b}^{\dagger}]=\mathbb{1}$, a few steps of algebra leave us with
\begin{align*}
\hat H_{\rm mol}&=\omega_g\mathbb{1} +\hat{\sigma}^{\dagger}\hat{\sigma}\left( \omega_e - \omega_g \right) +
\frac{1}{2}  \, \mu \, \nu_e^2 R_{eg}^2\hat{\sigma}^{\dagger}\hat{\sigma}\ + \nu_g(\hat{b}^{\dagger}\hat{b}+\frac{1}{2})- \frac{\, R_{eg} \, \mu \, \nu_e^2}{\sqrt{2\mu \nu_g}}(\hat{b}+ \hat{b}^{\dagger}) \, \hat{\sigma}^{\dagger}\hat{\sigma}
+ \frac{\left( \nu_e^2 - \nu_g^2 \right)}{4 \nu_g}(\hat{b}+ \hat{b}^{\dagger})^2 \hat{\sigma}^{\dagger}\hat{\sigma}
\end{align*}

From this, we identify the bare electronic transition frequency as $\omega_0=\omega_e-\omega_g+\mu \nu_e^2 R_{eg}^2/2$. Introducing the dimensionless coupling constants $\lambda_1= -R_{eg}\mu \nu_e^2/\sqrt{2\mu \nu_g^3}$ and $\lambda_2=(\nu_e^2-\nu_g^2)/4\nu_g^2$, we arrive at
\begin{equation}
    \hat H_{\rm mol}= \omega_g+\nu_g(\hat{b}^{\dagger}\hat{b}+\frac{1}{2})+ \omega_0\hat{\sigma}^{\dagger}\hat{\sigma} + \lambda_1 \nu_g(\hat{b}+ \hat{b}^{\dagger}) \hat{\sigma}^{\dagger}\hat{\sigma} + \lambda_2\nu_g (\hat{b}+ \hat{b}^{\dagger})^2\hat{\sigma}^{\dagger}\hat{\sigma}
\end{equation}
Finally, if we additionally assume that the curvature of the two oscillator potentials is approximately the same, $\nu_e \approx \nu_g$, then we can drop the $\lambda_2$-term, and $\hat H_{\rm mol}$ turns into the Hamiltonian \eqref{eq:HSO} used in the main text, with $\lambda \equiv \lambda_1$.

\end{document}